\documentclass[amssymb,prb,twocolumn,showpacs]{revtex4}

\usepackage{epsfig}
\usepackage{dcolumn}
\usepackage{amsmath}
\begin{document}

\title{Radiation-induced magnetoresistance oscillations with massive Dirac fermions.}

\author{Jes\'us I\~narrea$^{1,2}$ and Gloria Platero$^{2,3}$}

\address{$^1$Escuela Polit\'ecnica
Superior,Universidad Carlos III,Leganes,Madrid,28911 ,Spain\\
$^2$Instituto de Ciencia de Materiales, CSIC,
Cantoblanco,Madrid,28049,Spain.\\
$^3$Unidad Asociada al Instituto de Ciencia de Materiales, CSIC,
Cantoblanco,Madrid,28049,Spain.}


\begin{abstract}
We report on a theoretical study on the rise of radiation-induced magnetoresistance oscillations in two-dimensional
systems of massive Dirac fermions. We study the bilayer system of monolayer graphene and hexagonal
boron nitride (h-BN/graphene) and the trilayer system of
hexagonal boron nitride encapsulated graphene  (h-BN/graphene/h-BN).
We extend the radiation-driven electron orbit model that was previously devised to study
the same oscillations in
two-dimensional systems of Schr\"odinger electrons (GaAs/AlGaAS heterostructure) to the case of massive Dirac fermions.
In the simulations we obtain clear oscillations for radiation frequencies in
the terahertz and far-infrared bands.
We investigate also the power and temperatures dependence. For the former we obtain
similar results as for Schr\"odinger  electrons and predict the rise of zero resistance states. For the latter we obtain a similar qualitatively
dependence but quantitatively different when increasing
temperature. While in GaAs the oscillations are wiped out  in a few degrees,
interestingly enough, for massive Dirac fermions, we obtain observable
oscillations for temperatures above $100$ K and even at room temperature for the higher frequencies used in the
simulations.


\end{abstract}
\maketitle
\section{ Introduction}
Radiation-induced magnetoresistance ($R_{xx}$) oscillations (MIRO)\cite{mani1,zudov1} were unexpectedly discovered two
decades ago when a high mobility two dimensional electron gas (GaAs/AlGaAs heterostructure) under a vertical magnetic field ($B$) was irradiated
with microwaves (MW). Previously, an  influential and pioneering  theoretical work on these systems under similar conditions, (in a constant $B$
under radiation), had  been already carried out by  Ryzhii\cite{ryzhii} in the 70's. Both $B$ and the temperature ($T$) used in the experiment were very low, $T \sim  1K$ and $B< 1T$.
Along with this effect it was discovered also radiation-induced zero resistance states\cite{mani1,zudov1} (ZRS) when the radiation
power ($P$)  was sufficiently increased. These effects, which suggested a novel way of
radiation-matter interaction\cite{ina1,ina11,ina12},  received a lot of attention by the condensed matter community.
\begin{figure}
\centering \epsfxsize=3.5in \epsfysize=3.6in
\epsffile{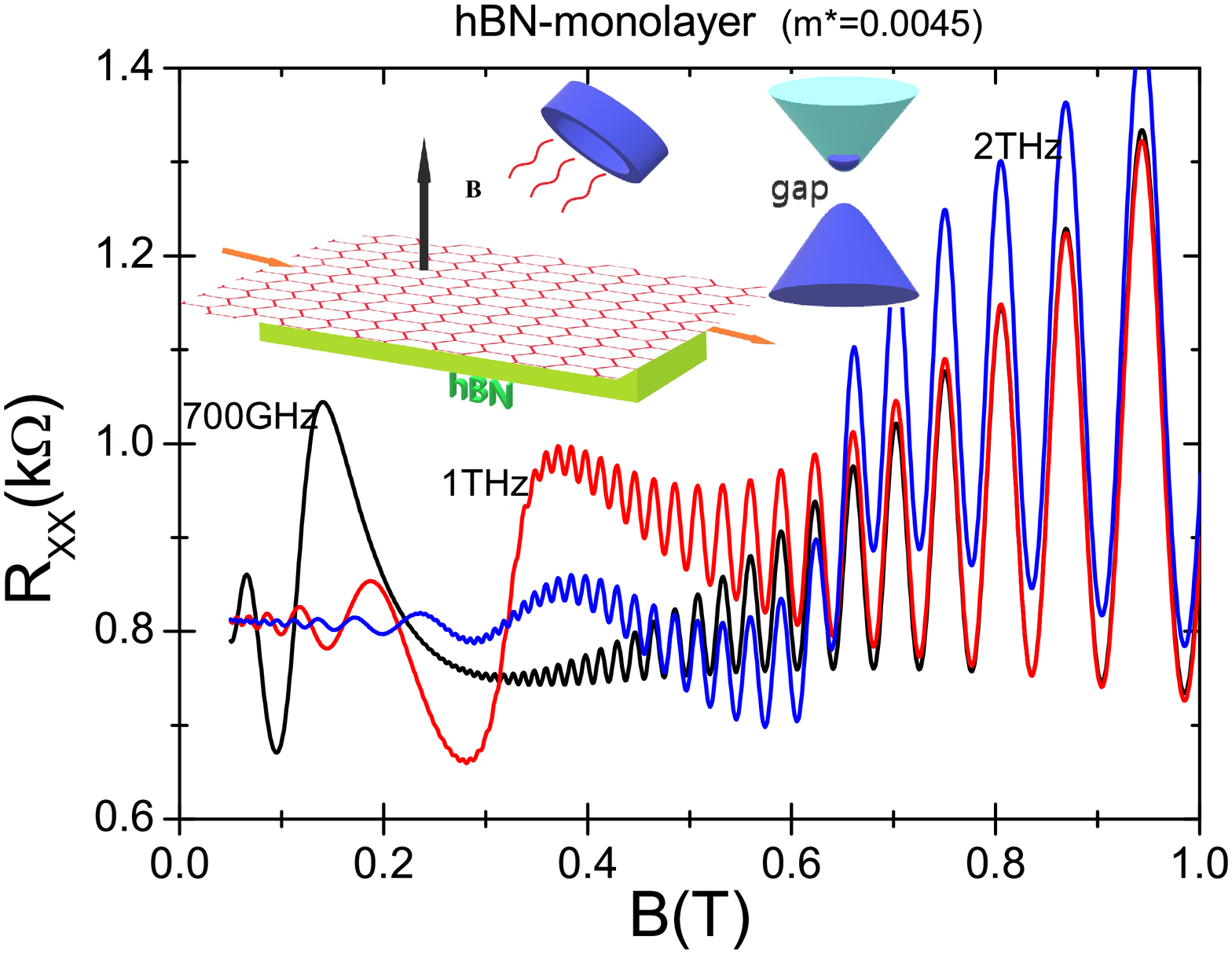}
\caption{Calculated  magnetoresistance $R_{xx}$ under radiation  vs  magnetic field for a bilayer of
h-BN and monolayer graphene. The three curves correspond to  frequencies of
700GHz, 1THz and 2THz. The oscillations amplitude gets smaller as  frequency increases (for $B<0.7$ T).
The inset displays a schematic diagram of the experimental setup basic features and the
band diagram of gapped monolayer graphene. $T=1$ K.
}
\end{figure}
 Thus,
 many experimental\cite{mani2,
mani3,willett,mani4,smet,yuan,mani5,wiedmann1,wiedmann2,kons1,vk,mani6,mani61,mani62,mani7,mani71,mani100} and
theoretical\cite{ina2,ina21,ina210,ina22,ina24,girvin,lei,rivera,ina41,ina42,ina5,ina51,ina61,inarashba,inaapl,ina5244,beltukov} works have  been carried out since their discovery.
In this way, they have been obtained in a wide range of two-dimensional (2D) platforms: GaAs/AlGaAs\cite{mani1,zudov1} and Ge/SiGe heterostructues
\cite{zudov2}, 2D electrons on liquid Helium \cite{chepe1} and in the system MgZnO/ZnO\cite{karcher}.
The MIRO obtained in all of them share the same features in terms of
frequency, power and temperature dependence. Besides, the oscillations minima are always $1/4$ cycle shifted irrespective of the platform and carrier.
Thus we can conclude that MIRO represents a universal effect and a novel demonstration of
radiation-matter interaction in 2D systems.

A natural extension of MIRO would be their study
on graphene\cite{novo} systems that we can consider paradigmatic in terms of 2D systems.
In graphene the carriers are massless Dirac fermions and in principle it is not
clear that they will couple with radiation to give MIRO in the same way as the Schr\"odinger electrons do.
Even in the case that MIRO could be obtained, it is not clear either if they would
keep all of the universal features that define MIRO or just a part of  them.
Nevertheless, it has been recently reported a remarkable theoretical  work that predicts
the appearance of radiation-induced resistance oscillations in monolayer and bilayer graphene\cite{manigraph}.
In the same way  an experimental work\cite{monch} has been published indicating
that MIRO can be obtained in h-BN encapsulated graphene keeping most of MIRO characteristics.

In this article, we focus on the presence of MIRO in gapped monolayer graphene systems. More specifically,
we study the bilayer system of monolayer graphene on top of hexagonal boron nitride (h-BN) and the
trilayer system of  h-BN
encapsulated monolayer graphene. In both systems the carriers are $massive$ Dirac fermions\cite{hunt,kinder,giovan,zollner,kibis,iurov}.
The latter is more in deep analyzed because it presents
one of the highest mobilities ($\mu \sim 3 \times 10^{5}$ $cm^{2}/Vs$) among the graphene systems and a high mobility is essential to
clearly observe MIRO. In our simulations we apply the theoretical model of {\it radiation-driven
electron orbits}\cite{ina2,ina21} successfully applied to Schr\"odinger electrons systems to address  MIRO and ZRS. For massive Dirac fermions
we recover magnetoresistance oscillations with an important difference in terms of
radiation frequency: according to our model,
these graphene systems are sensitive to terahertz (THz) and far-infrared frequencies instead of MW,
 giving rise to terahertz-induced resistance oscillations (TIRO). Nevertheless,  TIRO keep
similar  $P$ and $T$ dependence as MIRO. In the case of $P$ we obtain that as $P$ rises, the
oscillations amplitude rises as well following a sublinear law close to a square root dependence. We also  predict the rise of ZRS for
massive Dirac fermions when $P$ is high enough, as in MIRO. In the case of $T$,  we recover that oscillations decrease in
amplitude for increasing $T$. However, TIRO survive at much higher $T$ than the usual MIRO that are completely wiped out when increasing $T$ only a few degrees.
For TIRO we observe clear oscillations up to $100$ K for all simulations and even at room temperature ($300$ K) for the highest
frequencies used ($2.0 $ THz).

\begin{figure}
\centering \epsfxsize=3.7in \epsfysize=6.0in
\epsffile{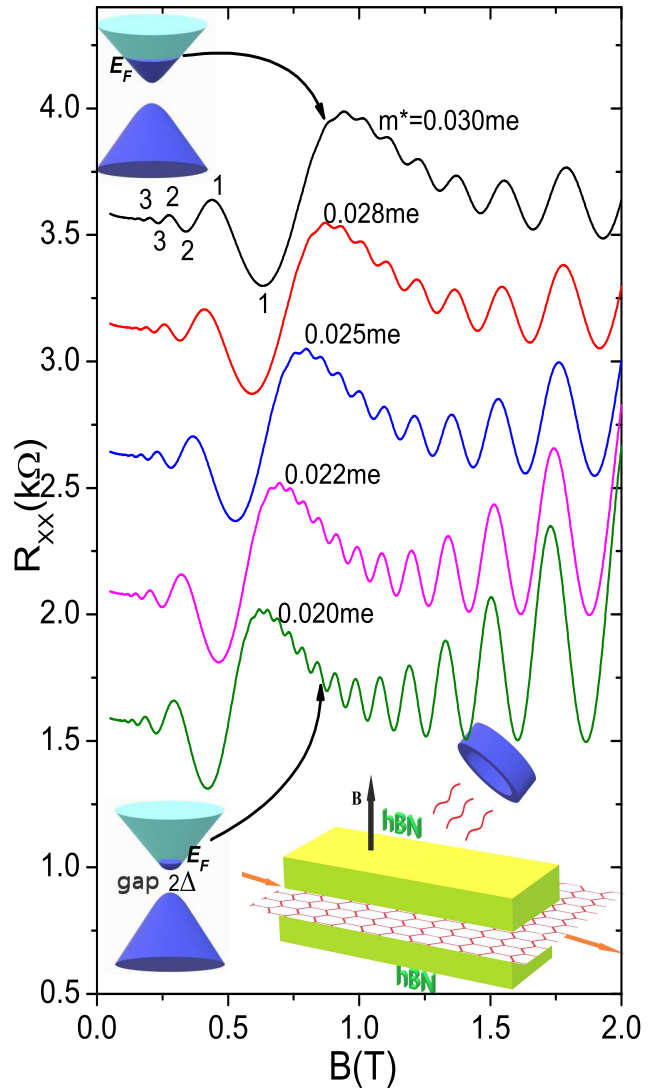}
\caption{Calculated magnetoresistance $R_{xx}$ under radiation  vs magnetic field for  h-BN
encapsulated monolayer graphene (hBN-monolayer graphene-hBN). The radiation
frequency is 700GHz. Five curves are exhibited corresponding to five different effective
masses or bandgaps. Every effective mass corresponds to a different bias between
the hBN layers (top and bottom). The inset shows a schematic diagram of the h-BN sandwiched monolayer
graphene  under a constant magnetic field and radiation. We present also two schematic band diagrams of gapped
graphene. The lower one corresponds to the lowest external bias and
 the lowest Fermi energy (green curve). The upper one
corresponds to the highest bias and a the highest
Fermi energy (black curve).    $T=1$ K.
}
\end{figure}
\begin{figure}
\centering \epsfxsize=3.5in \epsfysize=5.5in
\epsffile{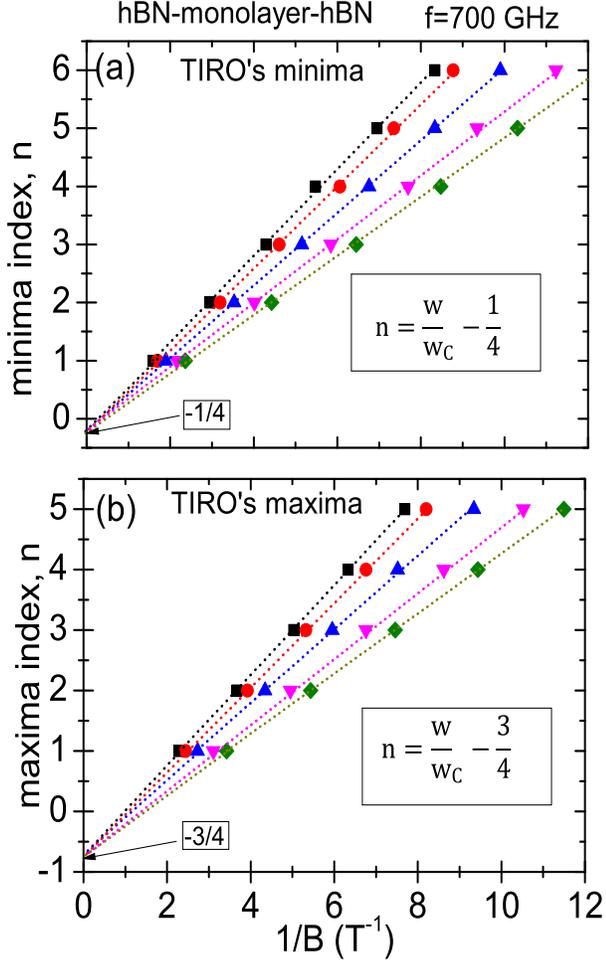}
\caption{TIRO's extrema vs the inverse of the magnetic field.
In the top panel we exhibit the minima index of Fig. 2 vs $1/B$. In the bottom panel we represent similar to the top panel for
the maxima index. As in MIRO, minima are 1/4 cycle shifted and maxima 3/4.
The dotted straight lines correspond to the minima and maxima fits. For the minima they
converge at -1/4 and for the maxima at -3/4.
This is expected from the theoretical model where the straight line equations are given by
 $n=\frac{w}{w_{c}}-\frac{1}{4}$ for the minima position and by $n=\frac{w}{w_{c}}-\frac{3}{4}$
for the maxima (see insets in both panels).
}
\end{figure}

\begin{figure}
\centering \epsfxsize=3.5in \epsfysize=5.5in
\epsffile{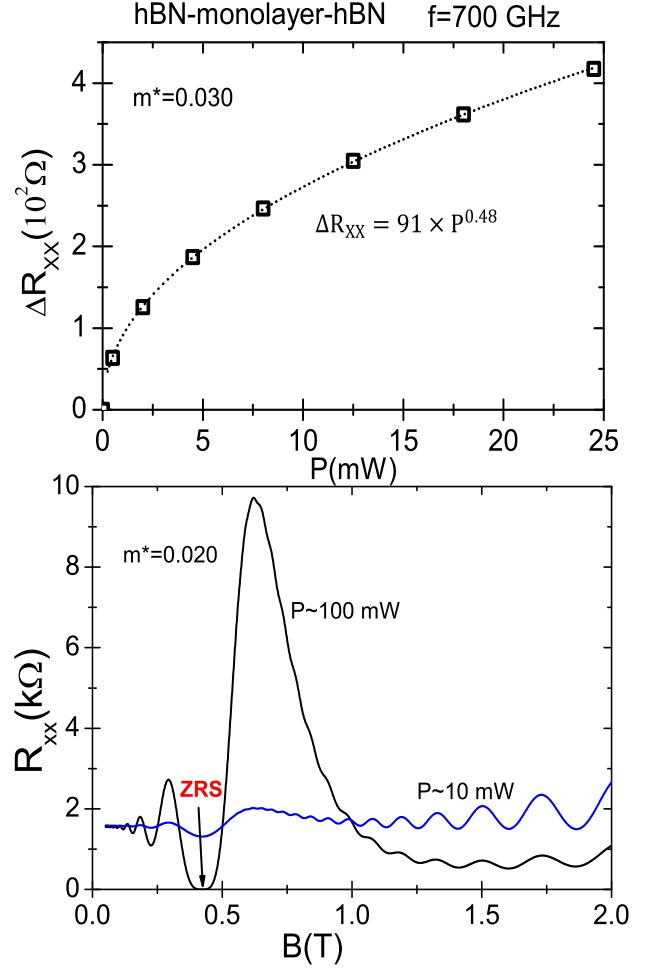}
\caption{Radiation power dependence of TIRO for a radiation frequency of 700 GHz. Upper  panel: Irradiated magnetoresistance amplitude of minimum 1 of $m^{*}=0.030m_{e}$
 curve in Fig. 2  vs radiation power. As in standard MIRO in semiconductor platforms,
 the calculated curve follows a sublinear relation (square root). In this case, $\Delta R_{xx}= 0.91 \times P^{0.48}$. Lower panel:
irradiated $R_{xx}$ vs $B$ for $m^{*}=0.020m_{e}$ case and two different powers, $P=10$ mW and  $P=100$ mW.
For the latter case we show that for a much higher radiation power it is
possible to reach a ZRS regime at $B\sim 0.4$ T. $T=1$ K.
}
\end{figure}
\begin{figure}
\centering \epsfxsize=3.5in \epsfysize=5.5in
\epsffile{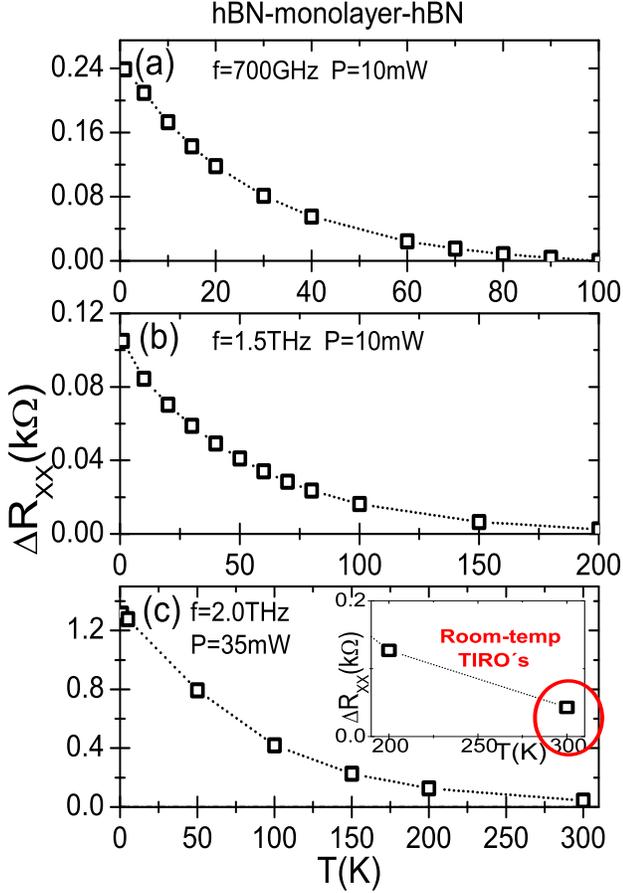}
\caption{Temperature dependence of the irradiated magnetoresistance
for frequencies of 700GHz in the upper panel, 1.5 THz in the middle
and 2 THz in the lower.
Radiation-induced oscillations persist at temperatures
as high as 100K in the upper panel and 200K in the middle panel.
In the lower one we present the striking result of observation of small TIRO at room T.
In the inset of the lower panel we exhibit a zoom-in of $T>200K$ region showing in detail the existence of
TIRO at room T
}
\end{figure}
\begin{figure}
\centering\epsfxsize=3.8in \epsfysize=3.in
\epsffile{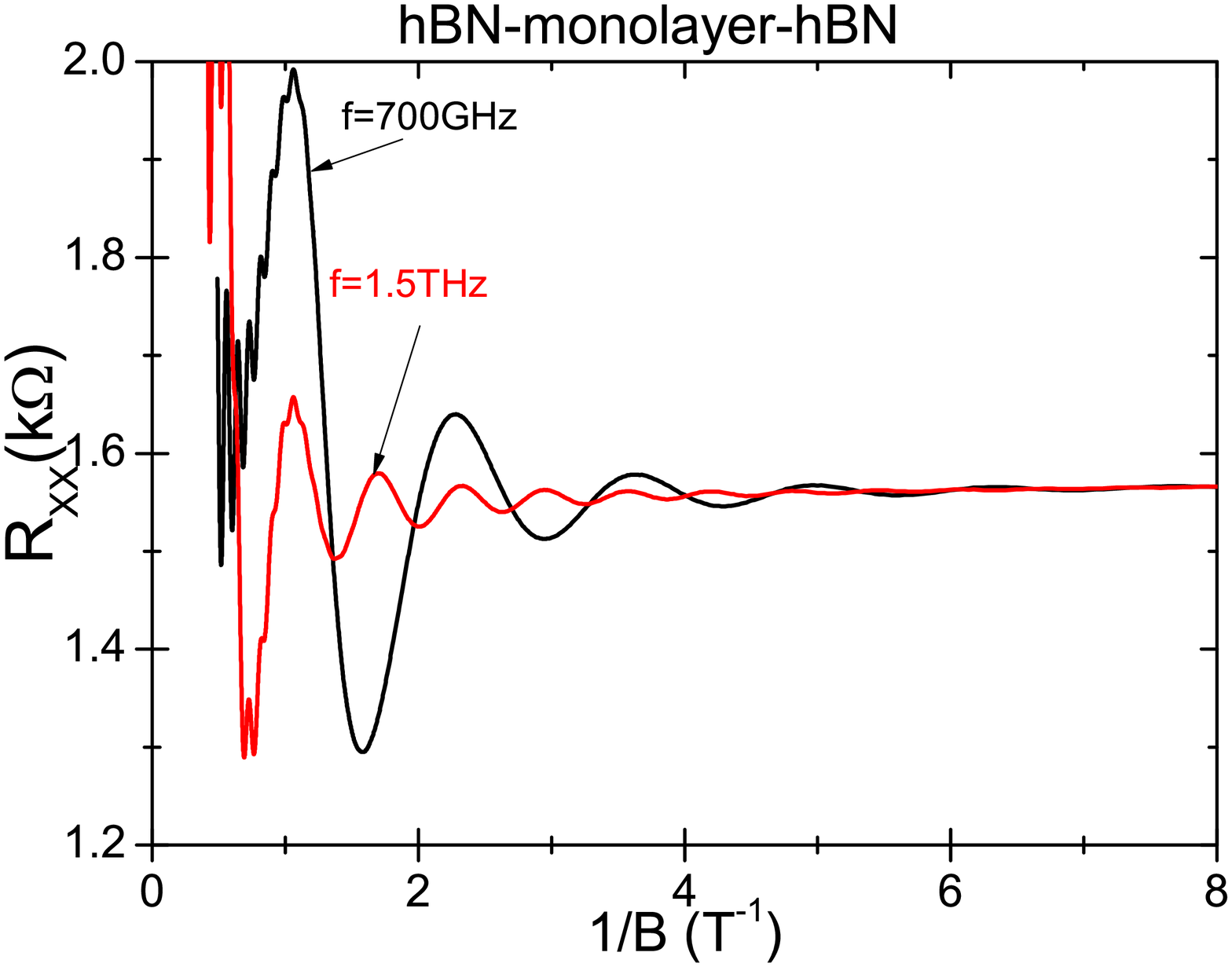}
\caption{ Frequency dependence of irradiated magnetoresistance vs the inverse
of the magnetic field. The exhibited frequencies are 700GHz and 1.5THz. We observe
that as  frequency increases the magnetoresistance oscillations amplitude get smaller.  $T=1$ K.}
\end{figure}

\section{ Theoretical Model}
Monolayer graphene is a 2D network of carbon atoms organized in
an hexagonal lattice that can be described as a triangular
Bravais structure with two atoms per unit cell denoted by A and B.
In the reciprocal lattice we obtain an hexagonal shaped Brillouin zone where
two out of the six corners are inequivalent. They are denoted as $K$ and $K'$.
When monolayer graphene is placed on top of a definite material (substrate), yielding a Van der Waals
heterostructure,   the interaction between
graphene  and the substrate gives rise to different potentials between the locations A and B, breaking
the carbon sublattice symmetry\cite{gusynin,gusynin2}.  As a result of this potential
asymmetry  a band gap opens at the Dirac point producing  gapped monolayer graphene (see insets of Figs. 1 and 2). Thus, monolayer graphene turns from
semimetal into a semiconductor, and remarkably enough,
the Dirac fermions become massive.

The Hamiltonian for the gapped monolayer graphene with a perpendicular magnetic field $\overrightarrow{B}=(0,0,B)$, for instance at the $K$ point, is given by:
\begin{equation}
H_{K}=\left( \begin{array} {cc}  \Delta   &   v_{F}\pi_{-} \\  v_{F}\pi_{+}  &   -\Delta  \end{array} \right)
\end{equation}
$v_{F}\simeq 1 \times 10^{6}$ m/s is the Fermi velocity, $\pi_{\pm} = \pi_{x}\pm i \pi_{y}$,
$\pi_{x}=P_{x}$
and $\pi_{y}=P_{y}+eBx$, where we have used the Landau gauge, $\overrightarrow{A}=(0,Bx,0)$.
$\Delta$ is a massive term that represents the potential asymmetry between the two sites.
The hamiltonian for the $K'$ valley is the same as for the $K$  but exchanging $\pi_{-}$ by $\pi_{+}$.
The corresponding eigenenergies and eigenfunctions can be readily calculated\cite{koshino,peeters}, for
the $K$ valley:
\begin{eqnarray}
E_{n,K}&=&\pm\sqrt{(\hbar w_{B})^{2} |n| +\Delta^{2}}; (n=\pm1,\pm2,....)\\
E_{0,K}&=&-\Delta  \nonumber\\
\end{eqnarray}
for the energies and
\begin{equation}
\Phi_{n,K} \propto \left( \begin{array}  {cc} \phi_{|n|-1}\left(x+\frac{\hbar k_{y}}{e B}\right) \\  \phi_{|n|}\left(x+\frac{\hbar k_{y}}{e B}\right)  \end{array} \right)
\end{equation}
for the eigenstates,
where  $\phi_{n}$ is the standard Landau level wave function (Landau state) and
$w_{B}=v_{F}\sqrt{2}/l_{B}$, $l_{B}$ being the magnetic length.\\
For the $K'$ valley:
\begin{eqnarray}
E_{n,K'}&=&\pm\sqrt{(\hbar w_{B})^{2} |n| +\Delta^{2}}; (n=\pm1,\pm2,....)\\
E_{0,K'}&=&\Delta   \nonumber\\
\end{eqnarray}
and
\begin{equation}
\Phi_{n,K'} \propto \left( \begin{array}  {cc} \phi_{|n|}\left(x+\frac{\hbar k_{y}}{e B}\right) \\  \phi_{|n|-1}\left(x+\frac{\hbar k_{y}}{e B}\right)  \end{array} \right)
\end{equation}

An important point of the gapped monolayer graphene  eigenstates  is that
at low lying energies in the conduction band, the amplitude of the wave function at the A site
is predominant\cite{koshino,peeters,cheng}, i.e., $|\phi_{|n|-1}| \gg |\phi_{|n|} |$ for the $K$ valley and
$|\phi_{|n|}| \gg |\phi_{|n-1|} |$ for the $K'$ valley. Accordingly, close to the conduction band
bottom we can approximately express the  eigenstates at the $K$ valley as,
\begin{equation}
\Phi_{n,K} \propto \left( \begin{array}  {cc} \phi_{|n|-1}\left(x+\frac{\hbar k_{y}}{e B}\right) \\  \sim 0  \end{array} \right)
\end{equation}
and a similar expression with $\phi_{|n|}\left(x+\frac{\hbar k_{y}}{e B}\right)$ for $\Phi_{n,K'}$.
Thus, close to the conduction band bottom in the gapped monolayer graphene the Dirac fermions are A-sublattice-polarised.
On the other hand, for eigenstates lying close to top of the valence band the amplitude of the
wave function is  predominant at the B sublattice and
then, they are B-sublattice polarised. The complete sublattice pseudospin polarisation is only
achieved right at the bottom (top) of the conduction (valence) band.
In the framework of the above approximation,  we can go
further and expand the hamiltonian near the conduction band bottom   and obtain
an effective expression for the Hamiltonians at the $K$ and $K'$ valleys\cite{koshino},
\begin{eqnarray}
H_{K}&\simeq& \frac{v^{2}}{2\Delta} \pi_{-} \pi_{+} = \frac{\pi^{2}}{2m^{\ast}}- \frac{\hbar w_{c}}{2}\\
H_{K'}&\simeq& \frac{v^{2}}{2\Delta} \pi_{+} \pi_{-} = \frac{\pi^{2}}{2m^{\ast}}+\frac{\hbar w_{c}}{2} \nonumber\\
\end{eqnarray}
The cyclotron frequency $w_{c}=eB/m^{\star}$,
$m^{\ast}$ being the effective mass of the corresponding {\it massive Dirac fermion},
\begin{equation}
m^{\ast}=\frac{\Delta}{v_{F}^{2}}
\end{equation}
Thus, $m^{\ast}$ turns out to be gap  dependent and can be tuned, for instance by an external bias or
by changing the substrate.
According to the above, for the eigenstates close to bottom of the conduction band and using
the effective expressions of the Hamiltonians  $H_{K/K'}$ we can write the
next equation for a stationary scenario,
\begin{equation}
\left[\frac{P_{x}^{2}}{2m^{\ast}}+\frac{(P_{y}^{2}+eBx)}{2m^{\ast}}\mp\frac{1}{2}\hbar w_{c}\right]\Phi_{n,K/K'} = E_{n,K/K'}\Phi_{n,K/K'}
\end{equation}
which is the Schr\"odinger equation in the presence of a static $B$. To obtain this equation we have taken into account that $\pi^{2}=\pi_{x}^{2}+\pi_{y}^{2}$ and the expressions
for  $\pi_{x}$ and $\pi_{y}$.
The $'-'$ sign would correspond to $K$ and $'+'$ to $K'$.

To address magnetotransport in gapped graphene in the presence of radiation  and $B$, we apply the {\it radiation-driven electron orbits model} that  was developed\cite{ina2,inarashba} to deal with MIRO and ZRS in high  mobility  2D
Schr\"odinger electrons system (GaAs/AlGaAs heterostructures). Its application to massive Dirac fermions  starts off from the above stationary equation.
The corresponding stationary  Hamiltonian can be turned into time-dependent including radiation (time-dependent) and thus,
\begin{eqnarray}
H(t)&=&\frac{P_{x}^{2}}{2m^{*}}+\frac{1}{2}m^{*}w_{c}^{2}(x-X)^{2}-eE_{dc}X +\nonumber \\
 & &+\frac{1}{2}m^{*}\frac{E_{dc}^{2}}{B^{2}}-eE_{0} x\cos wt
\end{eqnarray}
$E_{dc}$ is the driving DC electric field responsible for
the current, $X$ is the center of the Landau state orbit:
$X=-\frac{\hbar k_{y}}{eB}+\frac{eE_{dc}}{m^{*}w_{c}^{2}}$ and
$E_{0}$ the intensity of the radiation field.
$H(t)$ can be exactly solved allowing a solution for the massive Dirac fermion wave function
(for the $K$ valley):
\begin{equation}
\Phi_{n,K} \propto \left( \begin{array}  {cc} \phi_{|n-1|}\left(x+\frac{\hbar k_{y}}{e B}-x_{cl}(t),t)\right) \\  \sim 0 \end{array} \right)
\end{equation}
The time-dependent guiding center shift $x_{cl}(t)$ is given by:
\begin{eqnarray}
x_{cl}(t)&=&\frac{e^{- \gamma t/2} e E_{o}}{m^{*}\sqrt{(w_{c}^{2}-w^{2})^{2}+\gamma^{4}}}\sin wt\nonumber\\
&=& A(t)\sin wt
\end{eqnarray}
Then,  the wave function of $H(t)$ under radiation is
the same as in the dark  where the center is
displaced by $x_{cl}(t)$.
Therefore, according to this model  the
 radiation-driven Landau states, occupied by sublattice-polarised massive Dirac
 fermions, spatially and  harmonically  oscillate through the guiding center with the radiation frequency performing
 classical trajectories. This behaviour resembles the collective motion of electric charge. i.e., a plasmon-like mode, but in
 this case driven by radiation.

 In this swinging motion electrons interact with
 the lattice ions resulting in a damping process and giving rise to acoustic phonons. They can interact as well
 with different sources of disorder: defects such as graphene wrinkles and corrugations and with the sample edges yielding the damping of the plasmon-like mode.
The damping is phenomenologically introduced through the  $\gamma$-dependent damping term  in the previous $x_{cl}(t)$ equation.
Accordingly, the damping of the  Landau states-plasmon-like  motion
comes mainly from two sources. One is temperature independent where we include
disorder coming from graphene wrinkles and corrugations and the sample edges. The second is temperature dependent and comes
from the interactions of electrons with acoustic phonon modes (lattices ions).
Thus, we calculate $\gamma$ according to the phenomenological expression $\gamma=a +b(T)$,  $a$ being an average frequency term
representing the driven-Landau states oscillations and accordingly  $a \sim 10^{12}s^{-1}$, i.e., the parameter
$a$ is of the order of the oscillations frequency.  $b(T)$ is the
electron scattering rate with acoustic phonons
that depends linearly with $T$ \cite{sarma,sarma2,principi} according to $b(T)=1/\tau_{ac} \simeq (10^{11}-10^{12})\times T$ $s^{-1}$ for
monolayer graphene.

To study the magnetotransport in these systems and calculate $R_{xx}$ we consider
the long range Coulomb disorder (charged impurities) as the main source of scattering in graphene\cite{sarma}.
In this scenario an essential result of the model is that, under radiation,  the scattering process  of
electrons  with charged impurities turns out
to be dramatically  modified\cite{ina2,ina21,ina30,kerner,park}.
Thus, the key variable  is
the change of the guiding center coordinate ($\Delta X$) for the Landau
states involved in the scattering process\cite{miura}.
Under radiation $\Delta X$ turns into a harmonic function, increasing and
decreasing with the radiation frequency. When  $\Delta X$ increases, we obtain MIRO's peaks
in irradiated $R_{xx}$,
and when it lowers we obtain MIRO's valleys.
The final expression for the irradiated  $\Delta X$ is given by\cite{inahole,inarashba}:
\begin{equation}
\Delta X= \Delta X(0)-A  \sin \left(2\pi  \frac{ w}{w_{c}}\right)
\end{equation}
where $\Delta X(0)$ is the distance between the guiding centers of the final and initial
Landau states in the dark.

We use a semiclassical Botzmann theory
to calculate the longitudinal conductivity $\sigma_{xx}$\cite{titeica,ando,ridley,miura,sarma}:
\begin{equation}
\sigma_{xx}=2e^{2} \int_{0}^{\infty} dE \rho_{i}(E) (\Delta X)^{2}W_{I}\left( -\frac{df(E)}{dE}  \right)
\end{equation}
being $E$ the energy, $\rho_{i}(E)$ the density of
initial Landau states and $f(E)$ the Fermi distribution function.
$W_{I}$ is the scattering rate of
electrons with charged impurities that according to the Fermi's golden rule\cite{ridley,sarma,askerov}: $W_{I}=\frac{2\pi}{\hbar}|<\phi_{f}|V_{s}|\phi_{i}>|^{2}\delta(E_{i}-E_{f})$,
where $\phi_{i}$ and $\phi_{f}$ are the wave functions  corresponding to the initial and final Landau states respectively and
 $V_{s}$ is the scattering potential for charged impurities\cite{sarma,ando}. $E_{i}$ and $E_{f}$ are  the initial and final Landau states energies.
To obtain $R_{xx}$ we use the standard tensorial relation
$R_{xx}=\frac{\sigma_{xx}}{\sigma_{xx}^{2}+\sigma_{xy}^{2}}$, where
$\sigma_{xy}\simeq\frac{n_{i} e}{B}$, $n_{i}$ being the electrons density and $e$ the electron charge.

\section{Results}
In Fig. 1 we exhibit irradiated magnetoresistance vs $B$ for a bilayer heterostructure of
hBN and monolayer graphene. The three curves on the graph correspond to  frequencies of
700GHz, 1THz and 2THz and $T=1.0K$. The inset displays a schematic diagram of the experimental setup main features
and the band diagram of gapped monolayer graphene.
The electron effective mass used in the simulations is $m^{*}=0.0045 m_{e}$\cite{giovan} where $m_{e}$ is the bare electron mass.
Due to the low effective mass this system turns out to be sensitive to a much higher
radiation frequency than the usual MIRO. Thus, we obtain distinctive magnetoresistance oscillations
at frequencies clearly in the terahertz region and even close to the far infrared.  As in MIRO, we obtain in TIRO that the oscillations
become increasingly smaller as the frequency gets bigger keeping constant the
radiation power. Remarkably  enough, the oscillations are
mostly revealed at
very low magnetic fields, $B\leq 0.4 $ T.

In Fig. 2 we exhibit  irradiated magnetoresistance vs $B$ for a gated hBN encapsulated
monolayer graphene. The radiation frequency is $700 GHz$ and $T=1.0K$.
 The inset
shows the basic experimental set up: irradiated hBN sandwiched monolayer graphene under $B$.
We also exhibit  two schematic gapped graphene band diagrams at the lower and
the upper part of the figure.
The lower one corresponds to the lowest external applied bias. This means less injected electrons and
a  lower Fermi energy. The green  curve would represents this scenario. The upper band diagram
corresponds to the highest applied bias giving rise to more injected electrons and a higher
Fermi energy. This would correspond to  the black curve.
The hBN layers induce
a bandgap in the monolayer graphene that can be tuned by means of the vertical external bias between the two outer
hBN layers\cite{ruhe}. Thus,  we can select the Dirac fermions effective mass.
 Based on a previous work\cite{ruhe}, the external bias we have used in
the simulations ranges
from $0.40$ V/${\AA}$ to $1$ V/${\AA}$ that correspond to band gaps  from $230$ meV to $340$ meV.
Eventually, we have used five different effective masses, $m^{*}=\Delta/v_{F}^{2}$,  ranging from $0.020 m_{e}$ to $0.030m_{e}$.
Thus,  we present in Fig. 2 five  curves each one corresponding to a different  $m^{*}$.
 Likewise the case in Fig. 1, the current systems turns out
to be sensitive to Thz radiation and the oscillations turn up at low $B$ ($B\leq 1$ T).
As expected, when increasing the external bias (effective mass) the oscillations shift to higher $B$
and the number of oscillations increases too.

These new oscillations or TIRO fulfill the
basic characteristics of previous MIRO in semiconductor heterostructures.
A remarkable example is exhibited in Fig. 3 where we present the TIRO minima (upper panel) and maxima (lower panel) index vs the
inverse of $B$. The numerical values for both panels correspond to the $m^{*}=0.030m_{e}$ curve of Fig. 2.
We represent the fits for minima and maxima and the obtained dependence between the index $n$ and $1/B$ turns out to be an straight line.
For the minima the fits converge at an intercept of $-1/4$ and at $-3/4$ for the maxima.
Thus, as in MIRO\cite{inarashba,inahole}, minima $B$-positions are $1/4$ phase shifted and maxima are $3/4$ phase shifted.
These results are expected from the model. Thus, for the minima position it is obtained that, $n=\frac{w}{w_{c}}-\frac{1}{4}$
and for the maxima, $n=\frac{w}{w_{c}}-\frac{3}{4}$. These equations are in the insets.
This has been confirmed by recent experimental results\cite{monch}.

In Fig. 4 we exhibit the radiation power dependence of TIRO for hBN encapsulated graphene and a
frequency of $700$ GHz. In the upper  panel we present the irradiated
magnetoresistance amplitude of minimum 1 of the $m^{*}=0.030m_{e}$
 curve in Fig. 2  vs radiation power. As in standard MIRO in semiconductor platforms,
the calculated curve follows a sublinear relation, close to a square root function: $\Delta R_{xx} \propto \sqrt{P}$.
For our case we have obtained, $\Delta R_{xx}=91 \times P^{0.48}$ from the fit.
In the lower panel we present irradiated $R_{xx}$ vs $B$ for the $m^{*}=0.020m_{e}$ case and two different powers, $P=10$ mW and  $P=100$ mW.
Thus, for the latter case we predict that for a much  higher radiation power it would be
possible to reach a ZRS regime for irradiated massive Dirac fermions at $B \sim 0.4$ T for a frequency of $700$ GHz.

In Fig. 5 we present the temperature dependence of TIRO for hBN encapsulated monolayer graphene.
We exhibit three panels of the irradiated $R_{xx}$ amplitude vs $T$ for three different
frequencies, $700$ GHz (upper panel), $1.5$ THz (middle panel) and $2$ THz (lower panel). The exhibited amplitudes for the three of them
correspond to the minimum 1 of the $0.030m_{e}$ curve of Fig. 2. As in MIRO, all of them show that for increasing $T$
the TIRO amplitudes get smaller. For the upper panel we obtain visible TIRO up to around $100$ K, and for the middle one
we reach $200$ K. Interestingly enough, for the lower panel we obtain the striking result  that for the highest
frequency it would be possible to observe TIRO at room temperature ($300$ K). A zoom-in between $200$ K and $300$
K shows  clear, although small, oscillations at $300$ K.

In Fig. 6 we exhibit the frequency dependence of TIRO for hBN encapsulated graphene. We represent
two curves of irradiated $R_{xx}$ vs the inverse of $B$ for frequencies
 of $700$ GHz and $1.5$ THz. As in Fig. 1 we obtain much smaller TIRO for
 higher frequencies. Other simulations with bigger frequencies (not shown) have been run
 confirming this trend.

\section{Conclusions}
Summing up, we have presented a theoretical approach  on the rise of radiation-induced magnetoresistance oscillations in graphene systems
of massive Dirac fermions. We have considered the systems of monolayer graphene on top of
hBN and  the trilayer system of
gapped hBN encapsulated graphene. For both, Dirac fermions become massive.
We have applied  the radiation-driven electron orbit model formerly used
for 2D Schr\"odinger electron systems, adapted here to graphene systems.
In the simulations we have obtained clear oscillations that
show up for radiation frequencies in
the THz and far-infrared bands. This  contrasts with the two-dimensional Schr\"odinger electrons case,
that are mostly sensitive to microwave frequencies.
We have analyzed also the power, temperature and fequency dependence. For $P$ we have obtained
similar behaviour as in 2D Schr\"odinger electrons and  predicted
the rise of zero resistance states at high enough power. For $T$ we have obtained a
qualitatively similar behavior as standard MIRO in the sense that for higher temperature
oscillations amplitudes get smaller. However quantitatively speaking the behavior is different.
While in MIRO the oscillations are wiped out  in a few degrees,
 in TIRO we have obtained observable
oscillations for temperatures above $100$ K when using THz frequencies.
Finally, we want to highlight two important results of our model  from the applications standpoint:
first that TIRO show up in the terahertz band, very important frequency range from the
application perspective nowadays.  And secondly, TIRO  can be
obtained at room temperature. The two aspects can be of high  interest in the development
of novel optoelectronic devices for terahertz and infrared  frequencies such as sensors, phototransistors and
other graphene-based photovoltaic devices.


\section{Acknowledgments}
We thank R. G. Mani  for
stimulating discussions and a careful reading of our paper.
This work is supported by the MINECO (Spain) under grant MAT2017-86717-P and ITN Grant 234970 (EU).
Grupo de matematicas aplicadas a la materia condensada, (UC3M),
Unidad Asociada al CSIC.

\end{document}